# Scalability of On-chip Diffractive Optical Neural Networks

**SANAZ ZAREI**

*Department of Electrical Engineering, Sharif University of Technology, Tehran, Iran*
szarei@sharif.edu

**This short report focuses on the scalability challenges of the on-chip diffractive optical neural networks. It addresses an emerging gap in the literature, specifically around the limitations and challenges of scaling optical neural networks on a chip. A thorough investigation of diffractive optical neural networks provides evidence that such networks are not capable of performing complex tasks and exhibit significant performance degradation as the number of classification categories increases. Despite optimizations, these networks classify only 3-4 classes, suggesting fundamental limitations in their computational scale. The inherent scalability challenges in these systems are underscored by the fact that the design parameters, such as the number of diffractive layers, the number of neurons per layer, and the inter-layer distances, cannot substantially change the performance. Therefore, the on-chip diffraction-based approach provides a limited number of controllable degrees of freedom compared to electronic neural networks, restricting the complexity of functions an on-chip diffractive neural network can learn.**

**TOC Graphic:**

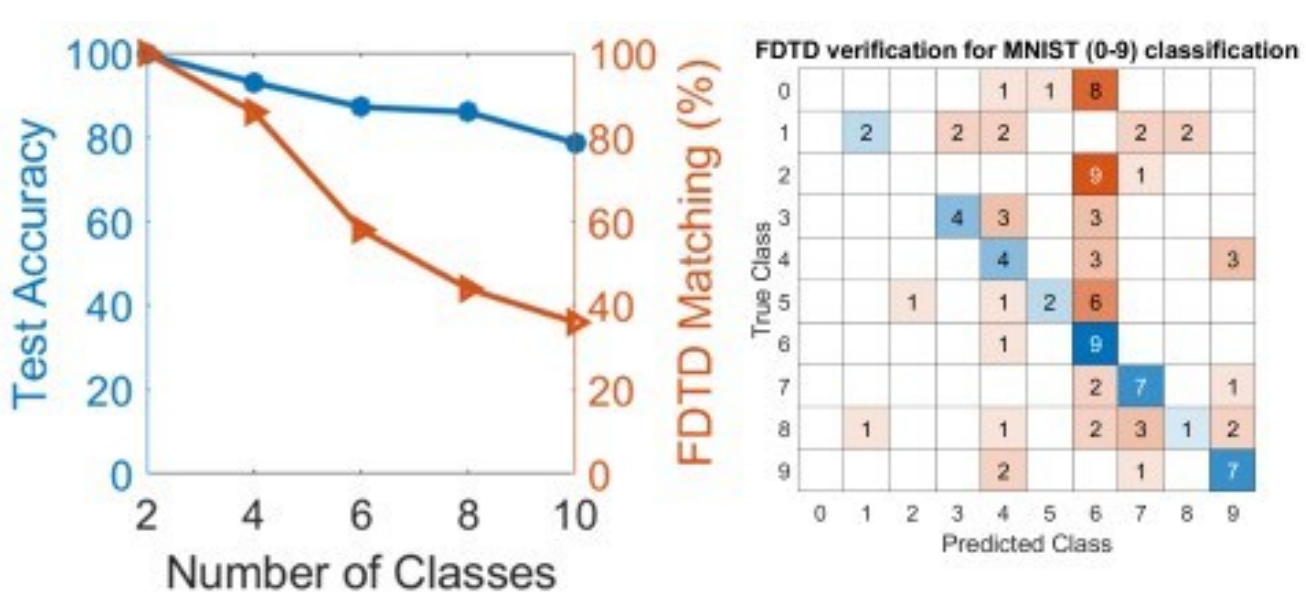

## 1. Introduction

Leveraged on high-contrast-transmit-array (HCTA) metasurfaces [1], several on-chip diffractive optical neural networks on a silicon-on-insulator (SOI) substrate have been demonstrated in previous works [2-10]. Despite many benefits offered by on-chip diffractive optical neural networks, like low-power consumption and light-speed parallel signal processing, challenges are faced because of deviations between diffraction-based analysis methods and experimental/full-wave electromagnetic verifications. While this discrepancy was mainly attributed to the limited capability of the diffraction-based analysis methods in modeling the evolution of optical fields through the network [10-11] and several previous works attempted to unravel the problem by applying a relatively large distance between successive metasurfaces to maintain stable interference [3-5], restricting multiple consecutive meta-atoms to be the same in the metasurfaces to decrease the mutual coupling between the adjacent meta-atoms [2-6], etc. [10], a theory is proposed in this article that the on-chip diffractive optical neural network has a very limited computational scale.

## 2. Modeling Approaches

Fourier optics methods are generally used to model optical field propagation through free space or across layers of materials. They are often applied to model diffraction, interference, and spatial frequency behavior of optical systems. In Fourier optics, the analysis is often centered around spatial frequencies, using the Fourier transform to describe diffraction in terms of wavevectors and wavefronts. This is a highly idealized and simplified view. On-chip diffractive optical neural networks often involve subwavelength features that can lead to complex phenomena like strong light confinement, localized modes, etc., which are difficult to model accurately by Fourier-optics methods, particularly when the wavelength is comparable to or larger than the feature size. At the cost of losing some physical details, however, Fourier-optics methods are often computationally very efficient because they reduce the problem to the frequency domain, leading to simpler algebraic equations that can be solved relatively quickly. On the other hand, full-wave solvers like FDTD offer a more accurate and comprehensive modeling approach and simulate the actual physical behavior of light by solving Maxwell's equations directly in time or frequency space. They account for all aspects of electromagnetic wave propagation, including nonlinear effects, material inhomogeneities, near-field effects, etc., though they are computationally more expensive. In the context of on-chip diffractive optical neural networks, the differences between the Fourier-optics analytical methods and full-wave electromagnetic solvers cause mismatches between their modeling results.

Many prior works attempted to reconcile the results obtained by the Fourier optics simulations and experimental/full-wave electromagnetic solvers [2-10]. Wang et al. considered two subwavelength slots to represents one phase shifter in their designed network, due to existing large phase contrasts between neighboring cells [2]. They also introduced a random phase offset with uniform distribution to their cells during the training stage to consider nanofabrication variations and measurement phase fluctuations [2]. They utilized the Rayleigh-Sommerfeld diffraction equation to model the wave propagation in the SOI slab waveguide. The Rayleigh-Sommerfeld diffraction formula is a more general form of Huygens' Principle, where the wave at any point is treated as being composed of contributions from every point on a wavefront, taking into account the distance from the observation point and the curvature of the wavefront. It is a frequency-domain approach and is often used in situations where near-field diffraction effects (such as spherical wavefronts) are important, however, it can be applied to both near-field and far-field diffraction. As it involves an integral over all the source points, it's not computationally very efficient. For X, Y, Z pattern recognition, Wang et al. achieved 98% numerical testing accuracy, 96% FDTD testing accuracy, and 92% experimental testing accuracy with 1550nm continuous wave input [2]. Fu et al. followed the Huygens-Fresnel principle in their work to model diffraction [3-4]. Their modeling method focused on how waves propagate over moderate distances (with the observation distance to be many wavelengths from the aperture ( $r \gg \lambda$ )) and required longer diffraction distances to reach a reasonable accuracy in predicting the device's behavior. In [3], in order to decrease the discrepancy between the results obtained by their Fourier optics method and FDTD full-wave solver, the authors assumed a large spacing between the adjacent hidden layers and utilized a slot group composed of three identical slots to approximate each neuron (cell). For the machine learning task of prediction of CHD from the UCI heart disease dataset, they could achieve 95.1%, 93.4%, and 91.8% matching between the two results for one-layer, two-layer, and three-layer neural network systems, respectively. In [4], they further applied an algorithm compensation method consisting of phase compensation and power compensation to reduce the impacts of the fabrication and measurement system errors. With compensation, they could improve the experimental classification accuracy of a one-layer and a three-layer Iris flower classifier from 56.7% and 60% to 86.7% and 90%, respectively. Also, an experimental classification accuracy of 86% under the external error compensation was achieved for the MNIST handwritten digits classifier. In [5], we exploited an integral of the two-dimensional Green's function of the scalar wave equation over all the source points to model the light propagation and diffraction inside the SOI slab waveguide. The presented integral can be regarded as a two-dimensional representation of the Huygens-Fresnel principle that can be applied in planar contexts. The on-chip diffractive optical neural network was trained with the aforementioned modeling method to function as a multifunctional optical logic gate, and a 100% matching score was achieved when verifying the numerical testing results by FDTD full-wave solvers. In our modeling method, while a large spacing between adjacent hidden layers was inevitable, a meta-atom consisting of just a single slot was sufficient to approximate a neuron. However, a meta-atom consisting of two identical slots could also end in an accurate response, but at the cost of a lower contrast ratio between the logical outputs. Yan et al. utilized a more subwavelength slot period with a larger subwavelength height in their work [6]. They chose the slot width (instead of the slot length) as the learnable parameter and set its variation range as [0, 100nm]. They also adopted a binary modulation such that the width of each slot was quantized to take values from {0,100} nm and set every three consecutive meta-atoms to be the same in their modeling to reduce the modulation error of the analytical model with respect to FDTD. Furthermore, they modeled the system error by including the Gaussian noise with a standard deviation of 0.3 to the trained phase and amplitude modulation coefficients during the evaluation. In their implementation, they founded their diffractive graph neural network upon integrated diffractive photonic computing units (DPUs) to generate the optical node features. The angular spectrum method was used in their work to analytically model the diffractive wave propagation. The angular spectrum method is based on the idea that a light wave passing through an optical system can be represented as a superposition of plane waves with various angles of propagation. These plane waves are propagated forward through the system by adjusting their phase based on the distance traveled. The angular spectrum method is particularly effective over short distances where the Fresnel approximation holds. However, for very short distances (distances comparable to the wavelength or less), the method may not be effective at capturing the extremely localized near-field effects that arise very close to the structure. For far-field modeling, padding that is the practice of artificially enlarging the spatial domain or the computational grid in the numerical simulation by adding extra zero-valued regions to the field distribution at the initial plane, can effectively improve the angular spectrum method's modeling ability [7]. Padding helps to both near-field and far-field modeling by improving the spatial and frequency resolution of the method, reducing the impact of numerical edge effects (produced by truncating the field at the boundaries of the computational grid), and reducing artifacts like aliasing. In [6], the authors could achieve comparable performances between the analytical model with system errors included and FDTD on synthetic SBM. They also reported analytical test accuracies higher than 86.5%, 74.4%, and 93.8% on Cora-ML, Citeseer, and Amazon Photo graph datasets. In an earlier work, we designed an MNIST digits classifier using the angular spectrum method [8]. My colleagues reported 91% matching between the Fourier optics and the FDTD full-wave solver. Liu et al. proposed a deep mapping regression model to characterize the process of light propagation in the on-chip diffractive layers [9]. They gathered a substantial amount of data and trained the deep mapping regression model to approximate the intricate Maxwell interactions within each hidden layer, and could improve the integration level of the on-chip diffractive optical neural network substantially. They optimized an ultra-compact one-layer and two-layer on-chip diffractive optical neural network and obtained matching scores of 96.5% and 96.6%, respectively, between numerical and FDTD testing for Iris flower classification. Sun et al. named several challenges in the advancement of on-chip diffractive optical neural networks such as large spacing between diffractive layers and open boundaries to ensure stable interference, decrease in the accuracy of analysis methods by increasing the number of diffractive layers, and the insufficiency of the analysis methods to analyze the evolution of loss of the optical field in the output ports [10]. They proposed a multimode on-chip diffractive optical neural network and utilized the eigenmodes as the neurons, for which the etching slots in a diffractive layer manipulated the coupling between them and realized their connection. The eigenmode analysis method was used to analyze the evolution of the optical field in the multimode on-chip diffractive optical neural network. They could design a more compact and energy-efficient multimode on-chip diffractive optical neural network with only one layer that showed 90% classification accuracy on Iris flower classification.

## 3. Theory

Throughout this article, the HCTA metasurface is a one-dimensional rectangular-shaped slot array with a lattice constant of 500nm that is etched in a silicon-on-insulator (SOI) substrate. The silicon top layer and buried oxide layer have thicknesses of 250nm and 2μm, respectively. A single neuron (or meta-atom) is formed by a single slot. The width and thickness of all slots (neurons) are fixed at 140nm and 250nm, respectively. For each neuron, the length of the slot is chosen as the learnable parameter. By altering the length of the slot between 100nm and 2.3μm, the transmission phase of a meta-atom (neuron) can be continuously tuned from 0-to-2π, while the transmission amplitude is near to 1 (please see Figure 1).

To further elaborate on the presented issue, the network performance on the classification of handwritten digits from the MNIST (Modified National Institute of Standards and Technology) dataset [12] is investigated. The scalability of the network can be studied by observing how the performance varies by increasing the number of digit classes. This can be a measure that to what extent the on-chip diffractive neural network can handle complex tasks.

Figure 2 illustrates the design parameters of the diffractive optical neural network trained as a digit classifier and the performance of the network versus the class number. The right axis represents the blind testing accuracy on the test dataset (calculated using the angular spectrum method [7, 13]), and the left axis indicates the matching percentage between angular spectrum numerical testing and Finite-Difference Time-Domain (FDTD) testing (conducted using the 2.5D variational solver of Lumerical Mode Solution). For FDTD testing, 100 handwritten digit images are randomly chosen from the test dataset images that were successfully classified by the angular spectrum method. As shown, the accuracy of the network decreases significantly as the number of digit classes increases. While for binary (0-1) digits classification, the numerical testing accuracy and the FDTD matching percentage are 99.71% and 100%, respectively, for ten (0-9) digits classification, the test accuracy drops to 78.67% and the FDTD matching percentage falls to 36%.

The classification performance of the diffractive optical neural network is further evaluated with respect to other design parameters such as the number of diffractive layers (metasurfaces), the number of meta-atoms (neurons) per metasurface (layer), the distance between two neighboring layers, and the distance between the last layer and the output layer (see Figure 3). These evaluations are performed for four handwritten digit (0-3) classes. As is evident in Figure 3(a), increasing the layers number results in higher numerical testing accuracy. However, beyond 3 layers, inferior FDTD matching is achieved. Also, due to Figure 3(b), increasing the number of neurons doesn't lead to a dramatic improvement in the testing accuracy, either numerical or FDTD testing. Figures 3(c) and 3(d) reveal the robustness of the diffractive optical neural network performance with respect to the distance between two neighboring layers and the distance between the last layer and the output layer, which does not bring about a further enhancement in testing accuracy.

The classification results presented in Figure 3 for four handwritten digits (0-3) classes are confirmed by the classification results introduced in Table 1 for ten handwritten digits (0-9) classes. The reported results imply that the classification accuracy of the diffractive optical neural network for ten handwritten digits (0-9) classes can't be advanced remarkably by increasing the number of layers, the number of neurons per layer, and the distance between layers.

## 4. Discussion

Attested by the investigations in Figure 2, the on-chip diffractive neural network is unable to learn more elaborate input/output relationships imposed by increasing the number of classes in the MNIST handwritten digits classification. Also, in contradiction to conventional neural networks and based on the presented results in Figure 3 and Table 1, the on-chip neural network cannot better fit more complex data/functions by adding more layers or more neurons per layer. These observations strengthen the theory that this architecture does not work well with such data. Furthermore, as perceived from Figure 3 and Table 1, altering the distance between the layers (and also the distance between the last layer and the output layer) does not provide a very effective means of control over the performance of the on-chip diffractive optical neural network.

The essential point that should be considered in Figure 2 is that the small distance of 17μm between neighboring layers raises 100% FDTD matching for the binary (0-1) digits classifier and only 36% FDTD matching for the ten (0-9) digits classifier. This depresses the necessity of choosing a relatively larger distance to achieve a better classification performance for this network and introduces task complexity as the most forceful factor.

Subwavelength metasurfaces cause strong diffraction, scattering, and interference effects in the near-field (close to them). Fourier-optics methods typically focus on far-field behavior, ignoring the strong near-field effects that become important in subwavelength metasurfaces. FDTD solvers model both near-field and far-field effects, leading to more accurate predictions. Therefore, measuring the electric field at a farther distance can reduce the mismatch between Fourier-optics and FDTD, especially for subwavelength metasurfaces. This is because at larger distances, the interaction between the light and the metasurface is less localized, and the near-field effects become less significant. However, if the metasurface causes significant diffraction or scattering (even at far distances), the far-field pattern can still be complex, and there might be a large mismatch between Fourier-optics and FDTD. As is demonstrated in Fig. 3(c), for a (0-3) digit classifier that is trained by the angular spectrum method [7, 13], distances as short as 7µm seem far enough such that the near-field effects are dissipated.

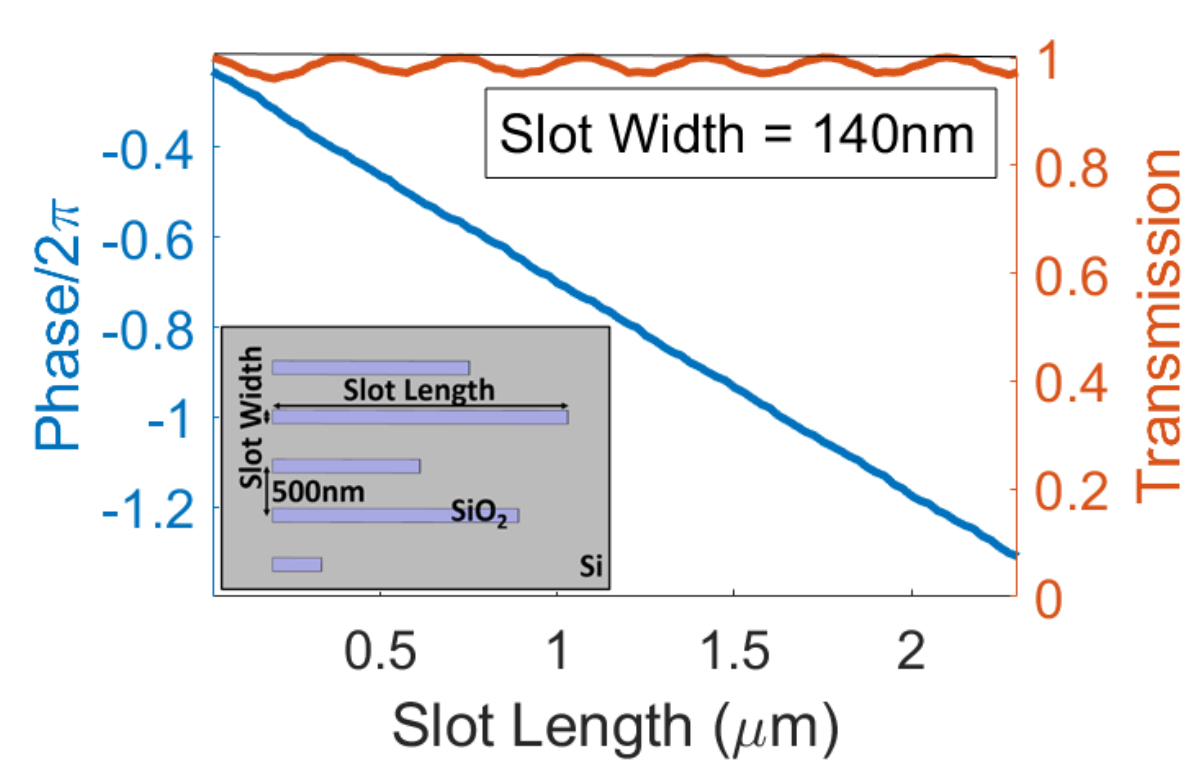

Figure 1. The variation of transmission phase and amplitude by slot length, fixing the slot width and height to 140nm and 250nm, respectively. The inset shows the 2D schematic of the HCTA metasurface consisting of subwavelength slots.

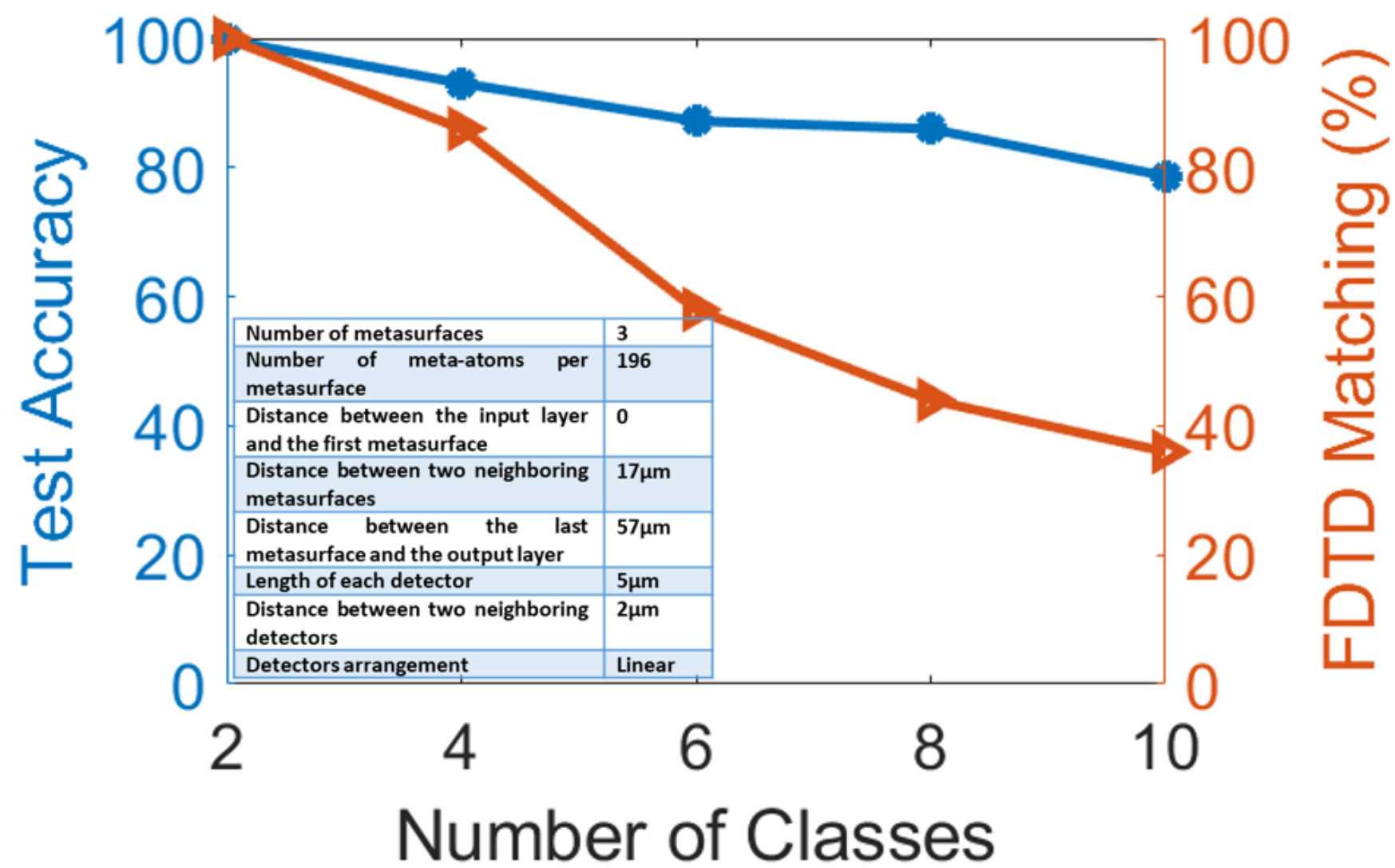

| | |
|---|---|
| Number of metasurfaces | 3 |
| Number of meta-atoms per metasurface | 196 |
| Distance between the input layer and the first metasurface | 0 |
| Distance between two neighboring metasurfaces | 17µm |
| Distance between the last metasurface and the output layer | 57µm |
| Length of each detector | 5µm |
| Distance between two neighboring detectors | 2µm |
| Detectors arrangement | Linear |

Figure 2. Blind testing accuracy of the diffractive optical neural network trained as a digit classifier with different numbers of digit classes. The right axis represents the accuracy computed by a diffraction-based analysis method, and the left axis depicts the matching percentage between the results achieved by the diffraction-based analysis method and FDTD verifications for 100 randomly selected handwritten digit images. The inset illustrates other design parameters of the digit classifiers.

Eventually, it is not indispensable to choose a large distance between successive layers in the on-chip diffractive optical neural networks to decrease the discrepancy between the diffraction-based analysis method and experimental/full-wave electromagnetic verifications. How far the diffractive layers should be placed strongly depends on the utilized diffraction-based analysis method. It is also affected by the extent of the complexity that the subwavelength metasurfaces introduce, such as localized modes, edge effects, strong diffraction, material inhomogeneities, and nonlinear effects. As an example, our previously presented multifunctional logic gate with its tremendous device footprint can be named [5]. It was designed using an integral of the two-dimensional Green's function of the scalar wave equation over all the source points, that needs the observation distance to be many wavelengths from the source ($r \gg \lambda$). Supplementary note 1 presents a more compact multifunctional logic gate that is designed by the angular spectrum method [7, 13].

A larger meta-atom made up of several identical smaller meta-atoms can increase the accuracy of the Fourier optics model because the metasurface can be treated as a more uniform periodic structure, and the model can be solved more efficiently. The primary downside of grouping multiple meta-atoms into a single one is losing some resolution regarding the individual interactions between meta-atoms, especially at smaller scales. In [5], for the multifunctional logic gates of the same size, meta-atoms made up of two identical slots result in a lower contrast ratio between the logical outputs compared to when meta-atoms are made up of one slot. In [14], for an MNIST digits classifier based on $Sb_2Se_3$-incorporated metasurfaces, super-meta-atoms of three identical meta-atoms didn't result in a better classification accuracy compared to when a sole meta-atom is regarded as the super-meta-atom. For this case, the number of super-meta-atoms was kept the same instead of the metasystem size.

If the slot length profiles of binary (0-1) digits classifier and ten (0-9) digits classifier are plotted in Figure 4, it can be inferred that such fluctuations in the slot length profiles of both classifiers similarly lead to comparable mutual interference effect between successive meta-atoms, which is expected to generate comparable level of error in the classification. But, in practice, one could achieve 100% FDTD accuracy, and the other could achieve only 36%. Therefore, the task complexity is playing a crucial role again, and the solution of choosing similar multiple meta-atoms as a super-meta-atom to reduce mutual interference seems to be unavailing in this case.

The confusion matrices for 4-, 6-, 8-, and 10-digit classifiers of Figure 2, calculated by the diffraction-based analysis method and FDTD, are depicted in the supplementary Figure S2. From the FDTD results in Figure S2, it can be deduced that all the classifiers can almost classify 3 or 4 digits appropriately, and the rest of the digits have not been properly learned. Besides, from Figures S2(f) and S2(h), the digits 0, 2, and 5 are frequently misclassified as digit 6. If the on-chip diffractive neural network is trained to perform as a four (0-2-5-6) digits classifier with similar design parameters to the ones in Figure 2, its blind testing accuracy and FDTD matching percentage reduce to 91.11% and 67%, respectively (compared to 93.04% and 86% for four (0-3) digits classifier). This indicates that the similarity between digits degrades the network performance to even lower than 3 properly-classified digits.

Also, the network performance on the classification of the Fashion MNIST dataset [12] versus the class number is studied in Table 2. The general observed trend is similar to the MNIST dataset, unless otherwise for the 2(0-1) class benchmark, which only shows 54% FDTD testing matching to the diffraction-based numerical testing. However, as is illustrated in Supplementary Figure S3, class #0 is misclassified by other classes in all the benchmarks for this dataset. This dramatically deteriorates the network performance in 2-class (0-1) classification, because of the high number of samples per class (50) in the FDTD testing of this benchmark. Complementarily, it can be understood from Supplementary Figure S3 that classes #1, 3, 5, and 7 are rather properly classified in all benchmarks. A 2-class Fashion MNIST benchmark on (1, 3) classes shows 89% matching percentage between FDTD testing and diffraction-based numerical testing (Supplementary Table S1). The matching percentage declines to 77% for the 4-class benchmark comprising (1,3,5,7) classes, which is very close to that of the 4-class (0-3) benchmark in Table 2. Furthermore, from Supplementary Table S1, it is clear that for the 10(0-9) class benchmark, changing the distance between two neighboring layers and the distance between the last layer and the output layer doesn't progress the matching between FDTD testing results and diffraction-based numerical testing results.

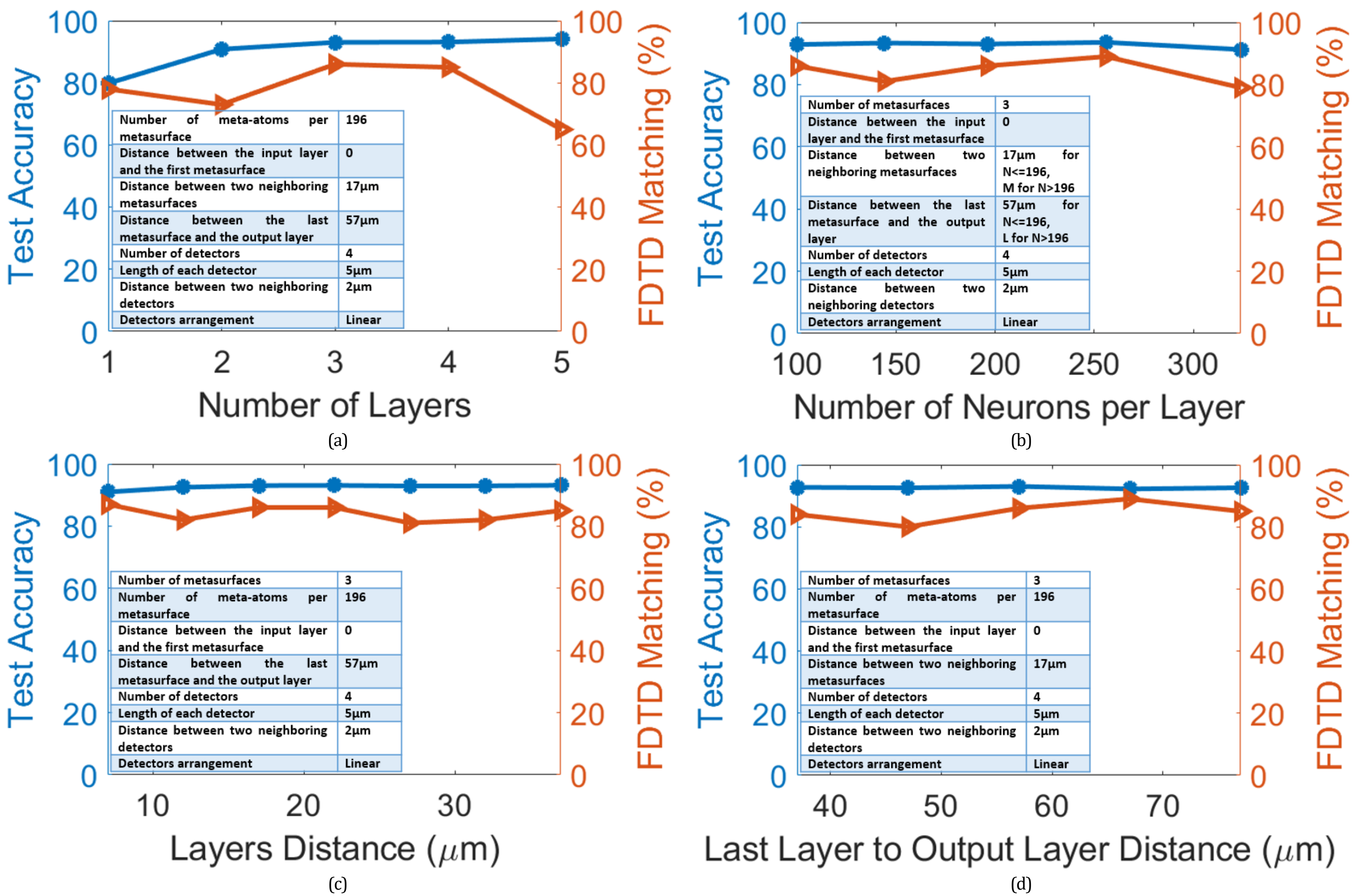

Figure 3. The variation of blind testing accuracy for the diffractive optical neural network trained as (0-3) digit classifier versus (a) the number of diffractive layers, (b) the number of neurons per layer, (c) the distance between two neighboring layers, and (d) the distance between the last layer and the output layer. The right axis presents the accuracy computed by a diffraction-based analysis method, and the left axis shows the matching percentage between the results achieved by the diffraction-based analysis method and FDTD verifications for 100 randomly selected handwritten digit images. The insets display all other design parameters of the networks.

**Table 1. Examples of the diffractive optical neural network trained as a (0-9) digit classifier**

| | Number of Layers | Number of neurons per layer | $D_{in}$ | $D_L$ | $D_{out}$ | Numerical testing accuracy | FDTD matching |
|---|---|---|---|---|---|---|---|
| 1 | 5 | 196 | 0 | 17μm | 57μm | 83.04% | 26% |
| 2 | 3 | 400 | 0 | 40μm | 120μm | 78.52% | 38% |
| 3 | 3 | 196 | 0 | 50μm | 50μm | 80.7% | 35% |

[a] $D_{in}$, $D_L$, and $D_{out}$ are the distance between the input layer and the first layer, the distance between two neighboring layers, and the distance between the last layer and the output layer, respectively.

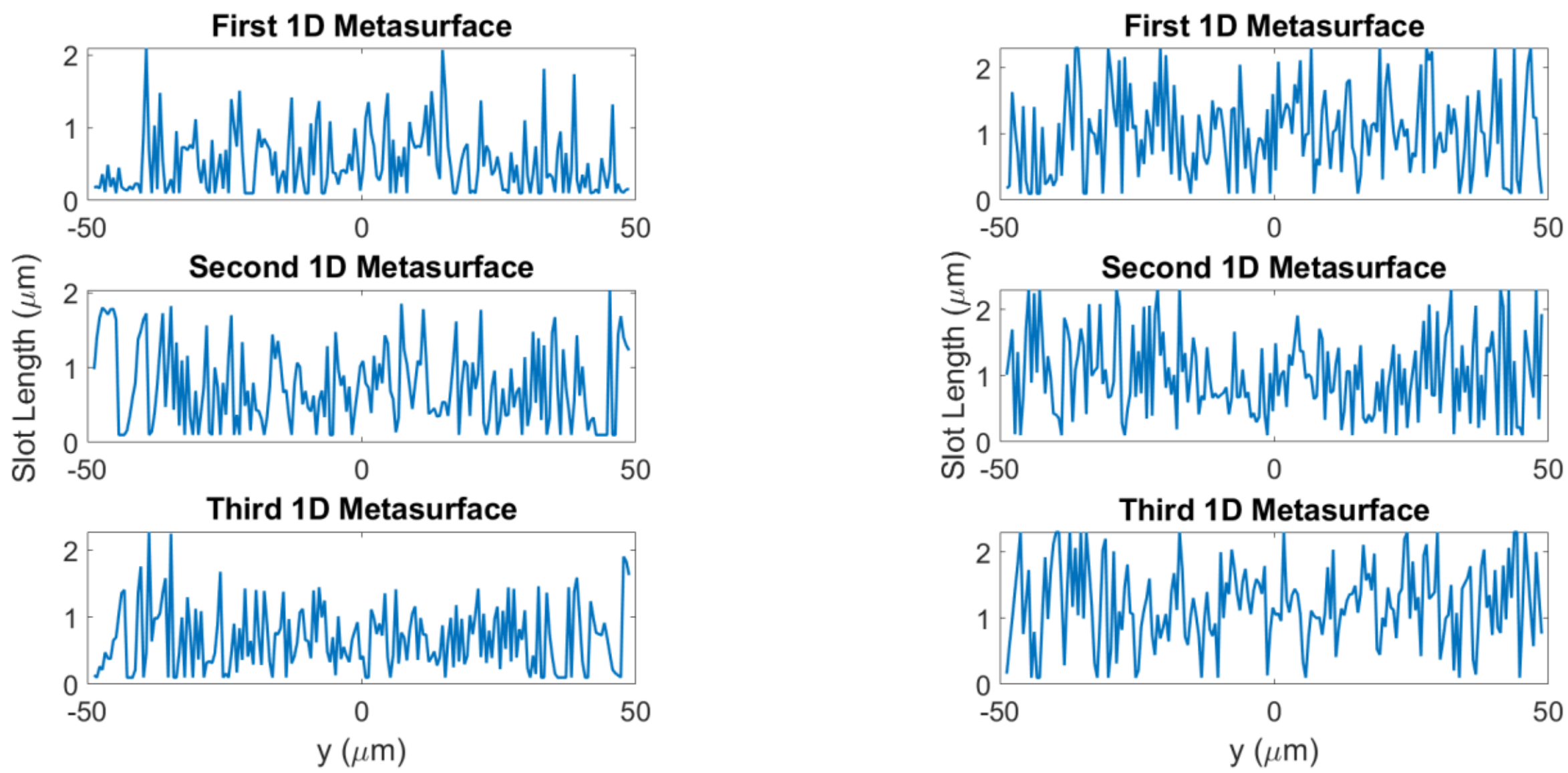

(a) (b)

Figure 4. The slot length profiles of the three-layer diffractive optical neural network trained as (a) a binary (0-1) digits classifier in Figure 2, (b) a ten (0-9) digits classifier in Figure 2.

**Table 2. The diffractive optical neural network trained for Fashion-MNIST classification with different numbers of classes**

| | Number of Classes | Number of Layers | Number of Neurons per Layer | $D_{in}$ | $D_L$ | $D_{out}$ | Numerical Testing Accuracy | FDTD Matching |
|---|---|---|---|---|---|---|---|---|
| 1 | 2(0-1) | 3 | 196 | 0 | 17μm | 57μm | 94.9% | 54% |
| 2 | 4(0-3) | 3 | 196 | 0 | 17μm | 57μm | 87.57% | 75% |
| 3 | 6(0-5) | 3 | 196 | 0 | 17μm | 57μm | 79.66% | 60% |
| 4 | 8(0-7) | 3 | 196 | 0 | 27μm | 57μm | 65.05% | 49% |
| 5 | 10(0-9) | 3 | 196 | 0 | 27μm | 57μm | 61.59% | 44% |

[a] $D_{in}$, $D_L$, and $D_{out}$ are the distance between the input layer and the first layer, the distance between two neighboring layers, and the distance between the last layer and the output layer, respectively.

## 5. Outlook and Future Works

In the context of on-chip diffractive optical neural networks, several strategies were exploited to reconcile the results achieved by Fourier-optics methods and the results achieved by FDTD. Differences in propagation models and discretization, however, raise discrepancies between Fourier-optics methods and FDTD. If the discrepancies between the two results are mainly because of the limited capability of the Fourier-optics methods in modeling the evolution of optical fields through the network, the in-situ training approach would dramatically improve the system performance. In this regard, the realization of programmable diffractive optical neural networks is of primary interest, and phase-change materials are ideal materials to realize programmability [14, 16-18]. Also, hybrid methods combining FDTD with Fourier-optics methods might provide a better trade-off between accuracy and computational efficiency. Since FDTD and Fourier-optics are both powerful methods, but with different strengths, a hybrid method where FDTD handles complex geometries and Fourier-optics is used for the diffraction propagation between layers in a simpler manner can be considered. For instance, it is possible to use FDTD to model the intricate details of the layers [9], while Fourier-optics is used for propagation between layers where the system is more homogeneous. In such situations, grid resolution matching or boundary condition adjustments between the methods might help to better resolve the mismatch. On the other hand, if the computational scale of the on-chip diffractive optical neural networks is fundamentally limited, other strategies like breaking down the task into smaller ones and then implementing the whole task by a system of integrated modules, each of which performing one smaller task [19-20], may be a feasible solution to take the advantage of the low-power consumption and light-speed parallel signal processing of the on-chip diffractive optical neural networks while circumventing the encountered restrictions.

This issue may not be limited to on-chip networks and may also be generalized to free-space diffractive optical neural networks. Although the current literature does not openly acknowledge this issue, examples of such inconsistencies (between numerical and experimental results) can be seen in several papers. For example, a research group at the University of Washington under the supervision of professor Arka Majumder reported that for classifying handwritten digit images, their single-layer diffractive optical neural network achieved an accuracy of 84.2% in numerical simulations, while in practical experiments obtained only an accuracy of 33% [21]. They attributed this low practical accuracy to the use of only a single diffractive layer in their optical neural network. Another research group in China, led by professor Xing Lin (inventor of free-space diffractive optical neural networks [22]), reported an accuracy of 97.6% for a three-layer optoelectronic diffractive deep neural network on the test data from the MNIST handwritten digit dataset, while a direct practical transfer of the trained model to the electro-optical system reduced the accuracy of the network to 63.9% [23]. They attributed this difference to non-idealities in the practical system that caused deviations from the computational model and the accumulation of computational errors in practice. However, it is not very unexpected if this was due to the limited computational scalability of the free-space networks as well.

It is worth mentioning that the computational scale of many optical neural network architectures like integrated Mach-Zehnder interferometer grids, micro-ring modulator arrays, etc., is typically limited to 4×4 matrix-vector multiplications or smaller [24]. Therefore, most existing optical neural networks still struggle with classic tasks and small datasets like MNIST and Fashion-MNIST [24]. As a result, the diffractive optical neural networks are also very likely to encounter such limitations in their computational scales.

## 6. Conclusion

In conclusion, as the complexity increases, the on-chip diffractive optical neural network based on HCTA metasurfaces encounters irresistible challenges of scalability. For the cases of MNIST handwritten digits and Fashion MNIST products classifications, this network can properly classify 3 or 4 classes, depending on the complexity and similarity of the classes it is trained to classify.

**Disclosures.** The author declares no competing financial interest.

**Data availability.** Data underlying the results presented in this study are not publicly available at this time but may be obtained from the authors upon reasonable request.

**Supporting Information:** Additional figures and tables as mentioned in the text.

**Supplementary information for**

# Scalability of On-chip Diffractive Optical Neural Networks

**SANAZ ZAREI**

Department of Electrical Engineering, Sharif University of Technology, Tehran, Iran
*szarei@sharif.edu

## This PDF file includes:

- **Supplementary Note 1: Compact Multifunctional Logic Gate**
- **Supplementary Figure S2: Confusion Matrices for MNIST Digits Classifiers**
- **Supplementary Figure S3: Confusion Matrices for Fashion MNIST Classifiers**
- **Supplementary Table S1 and S2: Complementary Benchmarks on Fashion MNIST Dataset**
- **References**

## Supplementary Note 1: Compact Multifunctional Logic Gate

While the design considerations that were illustrated in [1] resulted in a principally correct design, the overall footprint of the presented device became very large. If angular spectrum method [2-3] is utilized as the diffraction-based analysis method, realization of a multifunctional logic gate with a much smaller distance between the metalines is made feasible. Based on previous investigations, the restriction of similarity between multiple consecutive meta-atoms is also removed and each cell (super-meta-atom) is approximated with only one meta-atom. Analogous to the design presented in [1], this design is assumed to have five metalines (layers), with each metaline containing 100 meta-atoms (neurons). As each neuron is approximated with only one meta-atom (slot), the length of each metaline is 50μm. The input layer is located just before the first metaline and their distance is zero. The distance between two successive layers and the distance between the last layer (fifth metaline) and the output layer is set to only 25μm. The output layer of the network has two linearly-arranged detection regions which are representative of logic states "0" and "1". The length of each detection region is 2μm and the center-to-center distance between the two regions is 6μm. The metalines are regarded as parts of the propagation space (silicon slab waveguide) and no physical thickness should be devoted to them throughout the FDTD verifications. Similar input-output relationship as in [1] is used for training. The input combinations to the network, are the ones used in [1] resized by a factor of $1/2$ (because the length of metalines in this design are half of the ones in [1]). Figure S1(a) depicts the slot length profile of the designed multifunctional logic gate and Figures S1(b) and S1(c) show its performance. This new design has a footprint of 50μm×125μm (almost 24 times smaller than the previous design with a footprint of 100μm×1550μm). This footprint only involves the diffractive optical neural network, excluding the input waveguides and tapers.

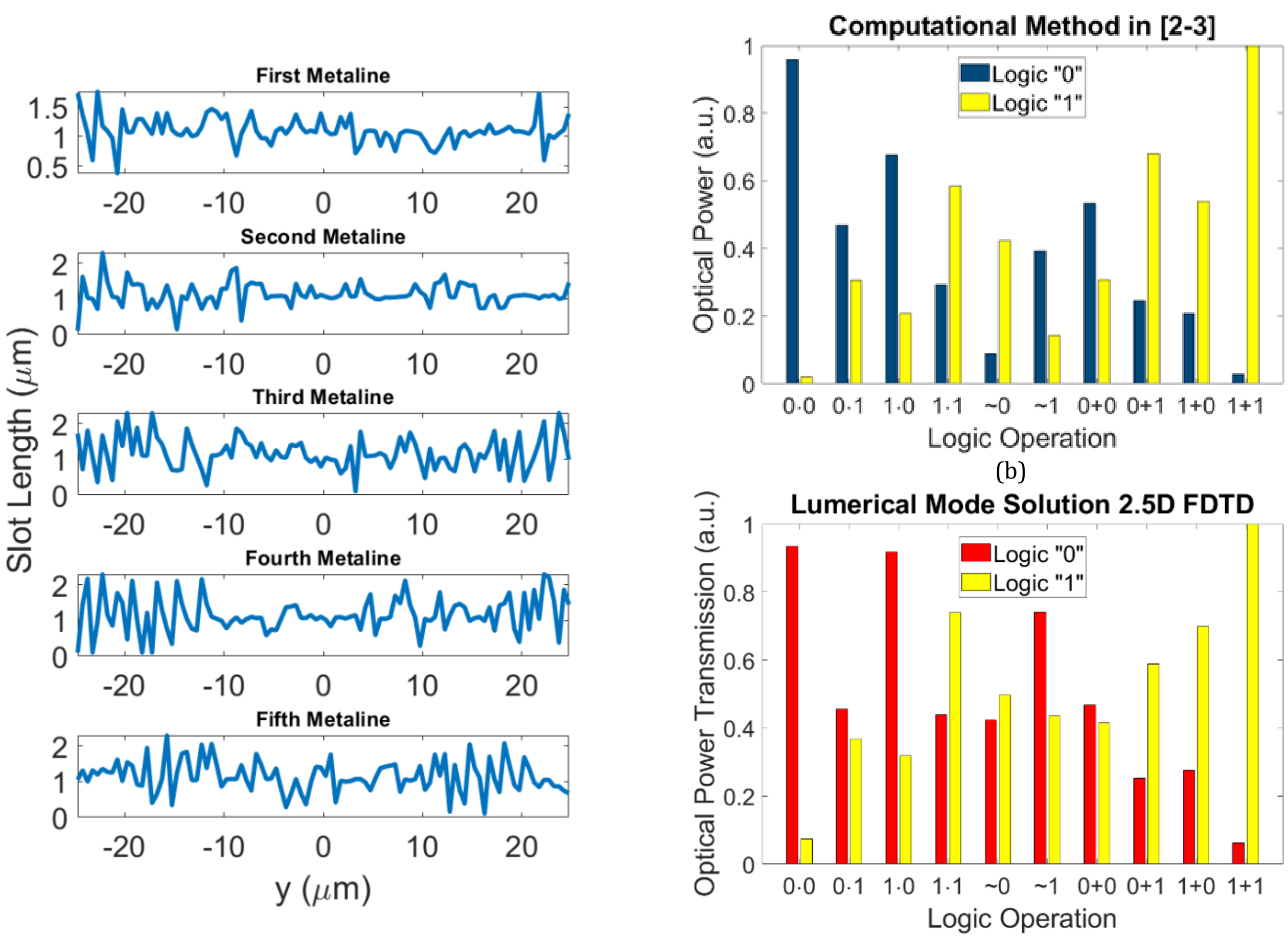

(a) (c)

Figure S1 (a) The slot length profile of the five-layer diffractive optical neural network trained as the multifunctional logic gate. The logic operation of the numerically-trained diffractive model, (b) numerical results, and (c) Lumerical Mode Solution 2.5D FDTD simulation results.

## Supplementary Figure S2: Confusion Matrices for MNIST Digits Classifiers

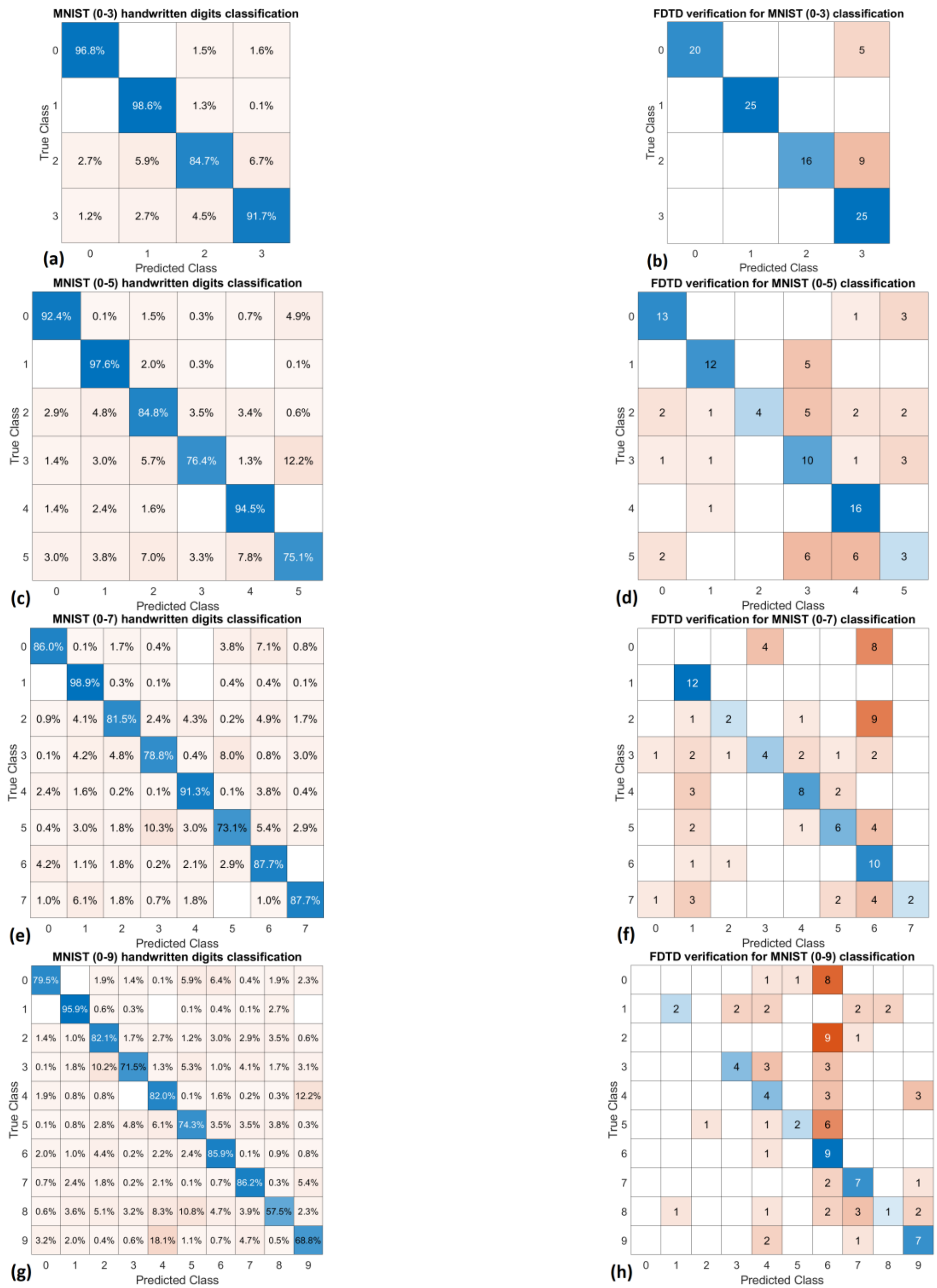

Figure S2. Row-normalized confusion matrix of the (a) four, (c) six, (e) eight, and (g) ten-digits classifier, calculated by the diffraction-based analysis method. Confusion matrix of the (b) four, (d) six, (f) eight, and (h) ten digits classifier, calculated by the 2.5D variational FDTD solver of Lumerical Mode Solution over 100 randomly-chosen handwritten digits images from the test data set images that were successfully classified by the diffraction-based analysis method.

## Supplementary Figure S3: Confusion Matrices for Fashion MNIST Classifiers

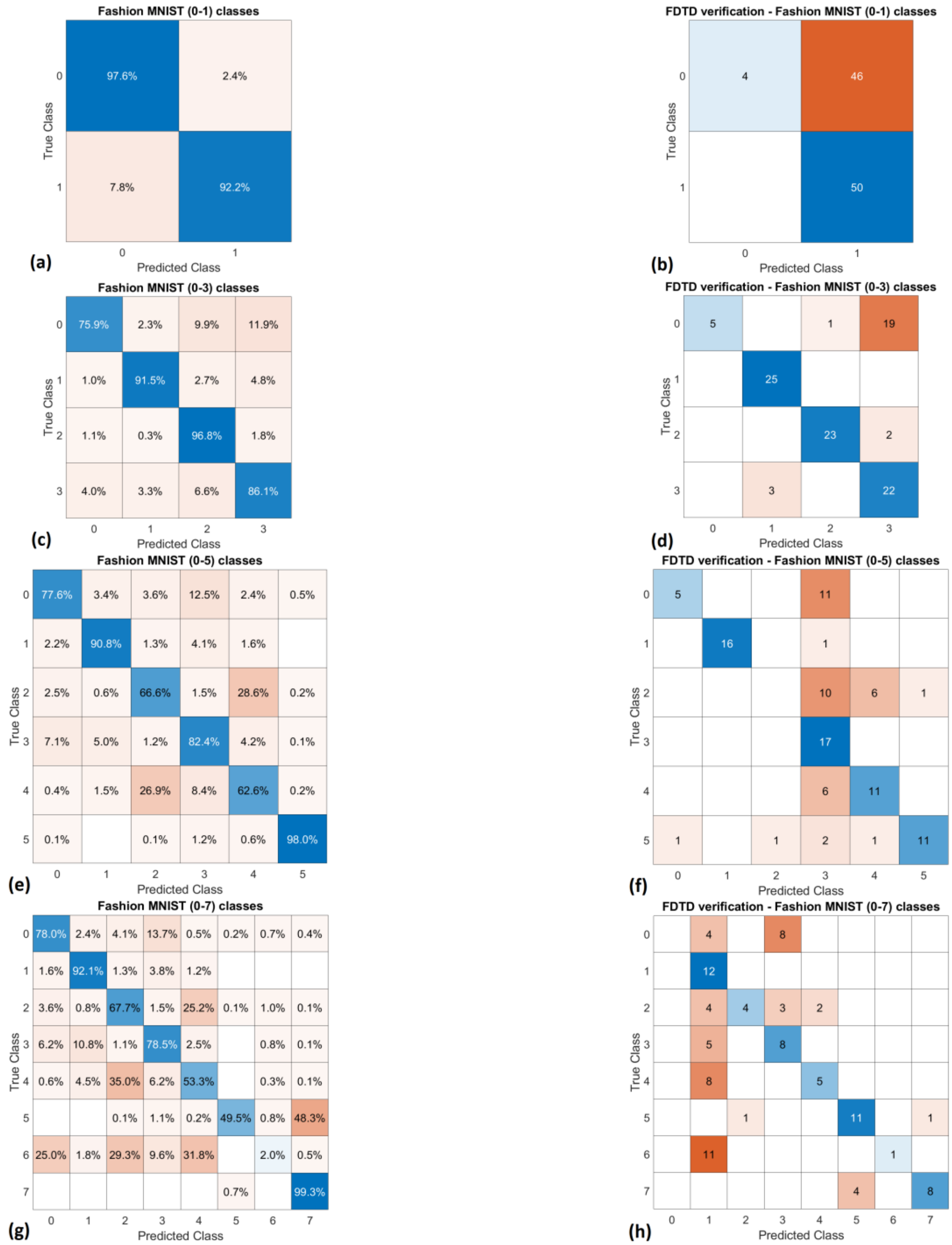

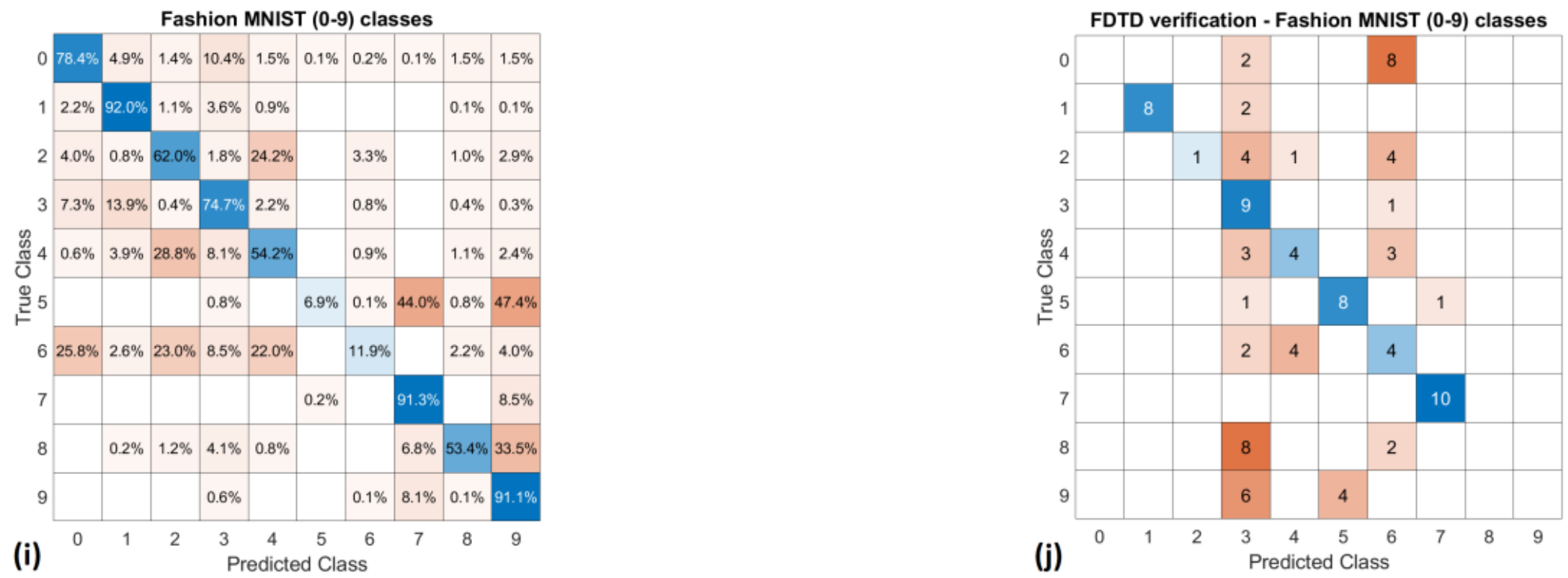

Figure S3. Row-normalized confusion matrix of the (a) two, (c) four, (e) six, (g) eight, and (i) ten-class Fashion MNIST classifier, calculated by the diffraction-based analysis method. Confusion matrix of the (b) two, (d) four, (f) six, (h) eight, and (j) ten classes classifier, calculated by the 2.5D variational FDTD solver of Lumerical Mode Solution over 100 randomly-chosen Fashion MNIST products images from the test data set images that were successfully classified by the diffraction-based analysis method.

## Supplementary Table S1 and S2: Complementary Benchmarks on Fashion MNIST Dataset

**Table S1. The diffractive optical neural network trained for Fashion-MNIST classification with different classes/number of classes**

| | Number of Classes | Number of Layers | Number of Neurons per Layer | $D_{in}$ | $D_L$ | $D_{out}$ | Numerical Testing Accuracy | FDTD Matching |
|---|---|---|---|---|---|---|---|---|
| 1 | 2(1,3) | 3 | 196 | 0 | 17μm | 57μm | 93.6% | 89% |
| 2 | 4(1,3,5,7) | 3 | 196 | 0 | 17μm | 57μm | 85.57% | 77% |
| 3 | 10(0-9) | 3 | 196 | 0 | 17μm | 57μm | 59.69% | 32% |
| 4 | 10(0-9) | 3 | 196 | 0 | 57um | 107um | 63.19% | 42% |

[a] $D_{in}$, $D_L$, and $D_{out}$ are the distance between the input layer and the first layer, the distance between two neighboring layers, and the distance between the last layer and the output layer, respectively.

**Table S2. Class names and labels in Fashion-MNIST dataset [4]**

| **Label** | 0 | 1 | 2 | 3 | 4 | 5 | 6 | 7 | 8 | 9 |
|---|---|---|---|---|---|---|---|---|---|---|
| **Class** | T-shirt/Top | Trouser | Pullover | Dress | Coat | Sandal | Shirt | Sneaker | Bag | Ankle boot |